\documentclass[conference]{IEEEtran}
\usepackage{cite}
\usepackage{amsmath,amssymb,amsfonts}
\usepackage{algorithmic}
\usepackage{graphicx}
\usepackage{textcomp}
\usepackage{xcolor}
\def\BibTeX{{\rm B\kern-.05em{\sc i\kern-.025em b}\kern-.08em
    T\kern-.1667em\lower.7ex\hbox{E}\kern-.125emX}}
    \usepackage{graphicx}
\usepackage{cite}
\usepackage{amsmath,amssymb,amsfonts}
\usepackage{algorithmic}
\usepackage{booktabs}

\usepackage{graphicx}
\usepackage{textcomp}
\usepackage{dcolumn}
\usepackage{istgame}
\usepackage{xcolor}
\usepackage{array}
\newcolumntype{P}[1]{>{\centering\arraybackslash}p{#1}}
\usepackage{forest}
\usepackage{tikz}
\usepackage{makecell}
\def\BibTeX{{\rm B\kern-.05em{\sc i\kern-.025em b}\kern-.08em
    T\kern-.1667em\lower.7ex\hbox{E}\kern-.125emX}}
\newcommand{\csp}[1]{{\color{blue}}}
\newcommand{\csg}[1]{}

\usepackage[strings]{underscore}
\usepackage[caption=false]{subfig}
\usepackage{caption}
 \usepackage[disable]{todonotes}
\usepackage[bottom=1.04in,top=0.75in,left=0.63in,right=0.63in]{geometry}
\usepackage{versions}
\excludeversion{longv}
\includeversion{shortv}
\usepackage{url}
\makeatletter
\newcommand{\@chapapp}{\relax}%
\makeatother
\begin{document}

\pagestyle{plain}

\title{Game-Theoretic Analysis of (Non-)Refundable Fees in
the Lightning Network}
\author{\IEEEauthorblockN{Satwik Prabhu Kumble}
\textit{TU Delft}
\and
\IEEEauthorblockN{Dick Epema}
\textit{TU Delft}
\and
\IEEEauthorblockN{Stefanie Roos}
\textit{University of Kaiserslautern-Landau}}


\maketitle

\begin{abstract}
In PCNs, nodes that forward payments between a source and a receiver are paid a small fee if the payment is successful. The fee is a compensation for temporarily committing funds to the payment. 
However, payments may fail due to insufficient funds or attacks, often after considerable delays of up to several days, leaving a node without compensation. Furthermore, attackers can intentionally cause failed payments, e.g., to infer private information (like channel balances), without any cost in fees. 
In this paper, we first use extensive form games to formally characterize the conditions that lead to rational intermediaries refusing (or agreeing) to forward payments. An intermediary’s decision to forward or not depends on the probability of failure, which they approximate based on past experience. We then propose and analyze an alternative fee model that allows the sender to determine and pay a fraction of the fee to intermediaries in a non-refundable manner. A rational sender chooses the fraction such that the intermediaries' utility for forwarding the payment exceeds their utility for not forwarding.
Our simulation study, based on real-world Lightning snapshots, confirms that our novel mechanism can increase the probability of successful payments by $12\%$ and decrease routing fees for senders by about $6\%$ if all nodes behave rationally. Furthermore, previously cost-free probing attacks now require that the attacker pays 1500 satoshis for every 1 million satoshis inferred. Finally, we propose a modification to the Hash Time Locked Contract to enable secure payments of the non-refundable fees. 
\end{abstract}

\begin{IEEEkeywords}
Payment Channel Networks, Extensive Form Games, Payment Fees, Probing attacks
\end{IEEEkeywords}

\section{Introduction}

Payment channel networks have emerged as a predominant solution to the scalability issues~\cite{gervais2016security} of Bitcoin~\cite{nakamoto2012bitcoin} and Ethereum~\cite{Buterin2013}.
They achieve better scalability, latency, and energy efficiency by offloading transactions between two users from the blockchain to an off-chain layer. 
Lightning, a payment channel network on top of Bitcoin with a liquidity of more than 100 million US dollar, has recently been adopted by some companies for real-world and real-time payments~\footnote{\tiny{https://bitcoinmagazine.com/culture/how-lightning-network-events-bring-adoption}}.




Multi-hop payments in Lightning are conducted using a two-phase protocol. During this two-phase protocol, senders and intermediaries are required to lock coins that cannot be used for other payments. Thus, intermediaries may be unable to start payments on their own or participate in other payments, meaning there is a collateral cost associated with locking. The cost depends on the amount of coins locked and the duration for which the coins are locked.
The sender pays intermediaries a small fee for locking coins. However, as the fee is only paid upon success, it is not guaranteed that the fee is a sufficient compensation. 

Consequently, it  can  be  a  desirable choice for rational intermediaries to not participate in payments if payments fail frequently. Failed payments can be a result of attacks. For instance, during \textit{probing attacks}~\cite{probing2022Nisslmueller,probing2019herrera,probing2020gijs,probing2020tikhomorov,Biryukov2021probing}, malicious parties attempt to infer the distribution of coins in one or more channels by repeatedly initiating payments via the channel(s) and then intentionally failing them after getting  feedback whether on they are possible. 


Our contribution in this paper is two-fold: First, we game-theoretically analyze the existing incentives for intermediaries in Lightning to participate in payments. We assume rational intermediaries who only lock if their expected utility, that is the fee gained minus the collateral cost, exceeds 0. They estimate their expected fee based on previous observations of payment success probability. We determine a \emph{Bayesian Nash Equilibrium}, where all parties act consistently with their beliefs to maximize their expected gain. 
Second, we modify the incentives such that partial non-refundable fees (smaller than the routing fee) are paid to intermediaries who lock, even if a payment fails. The complete routing fee is paid for payments that succeed.  We analyze the modified incentives game-theoretically and determine the respective Bayesian Nash Equilibrium. 
A payment sender has two options for choosing the partial fee. It can choose the partial fees such that intermediaries always have a positive utility, even for failed payments, guaranteeing that intermediaries always lock if they have sufficient funds. 
Alternatively, it can also aim to maximize its utility by paying a lower partial fee such that intermediaries are expected to have a positive utility but may not, due to inaccurate beliefs. In this case, intermediaries may refuse to lock. Our analysis is applicable to any source routing payment mechanism that uses a two-phase protocol based on locking and unlocking like Lightning.

We evaluate the concrete effect of our modified incentives on rational players in a simulation study. We find that these modified incentives increase the success ratio of payments by up to 12\%. The average routing fees for senders also decreases by up to 6\% as in the original Lightning model, recent payment failures of cheap paths force senders to choose more expensive paths. Our modified incentives model avoids this to some degree as intermediaries of cheaper paths have a higher incentive to participate in payments. Moreover, we find that our modified incentives result in a notable cost for probing attacks. If the balance in one direction of a channel is $n$ million satoshis, we find that the attacker requires approximately $1500n$ satoshis in fees to infer the balance. 
With over 4000 Bitcoin~\footnote{\tiny{https://1ml.com/statistics}} locked in Lightning, a widespread attack when applied to the complete network amounts to tens of thousands  of US dollars in cost.

\section{Background}
\label{sec:background}

In this section, we introduce the concepts of payment channel networks and game theory required in our analysis of two routing approaches.   

\subsection{Payment Channel Networks}\label{secpcn}
A payment channel~\cite{decker2015fast} enables two parties to conduct local payments between each other without the need of verifying every payment on-chain while ensuring that both parties can claim the funds that rightfully belong to them. For this, they deposit coins in a shared account. They can then use these coins for local payments. Each payment updates the distribution of coins between the two parties. This distribution is considered private. 
The blockchain acts as a mediator when the two parties disagree on the distribution.

Payment channel networks enable \emph{multi-hop payments} between users that do not share a payment channel network between them without requiring interaction with the blockchain. In such cases, the payment is routed via a path of existing payment channels from a source to a destination~\cite{decker2015fast}.
The Lightning Network (LN)~\cite{lightning} is the payment channel network on top of Bitcoin with more than 15,000 users~\footnotemark[2] .  Multi-hop payments in Lightning are conducted using a source-routed two-phase protocol.   \csp{changed} The receiver generates a payment secret and provides its hash $R$ to the sender. The payment amount is locked in each channel (with information on the hash) from the sender to the receiver. Then, the receiver provides the payment secret to its predecessor. The locked amount in each channel is released in the reverse order from the receiver to the sender once each party receives the payment secret from its successor. In this manner, payments are only finalized once it is known that the payment does reach the receiver. Senders compensate the intermediaries with a small fee if the payment is successful. 

\subsection{Route Finding Protocol}\label{secroute}
In Lightning, all channels are broadcast upon construction and hence the sender has a topology snapshot including the total capacities of all channels as well as the fee and time lock parameters. The sender can use this information to find paths to receivers that have low fees, low timelocks and high probability of success. Dijkstra's algorithm is used to find paths based a custom cost function. The cost function depends on fees, timelocks and success probabilities of the channels in the path. Let $(v_0,v_1,\ldots,v_n)$ be a path with $v_0$ being the source (or payer) and $v_n$ being the destination (or payee) found using \emph{LND}\footnote{\tiny{https://github.com/lightningnetwork/lnd}} (the predominant routing client used by Lightning users) and let $\alpha$ be the amount sent to $v_n$. Let $t_i$ be the timelock of channel $[v_i,v_{i+1}]$ and $\alpha_i$ be the amount being locked in the channel. The fee charged by intermediary $v_i$ is $b_i+\alpha_i\cdot \delta_i$, where $b_i$ is the \emph{base fee} and $\delta_i$ is the \emph{fee rate}. Both values are public and assigned during channel creation. The cheapest path is always constructed from the recipient to the sender. When $v_i$ is being considered to be added to the path, the cheapest sub-path from $v_{i+1}$ to $v_n$ has already been found. Then, the weight of the channel is defined as $weight_i=\alpha_i\cdot t_i\cdot r + f_i $. Here $r$ is a fixed risk factor  and $f_i$ is the fee paid by the sender to $v_i$. Here $\alpha_i\cdot t_i\cdot r$ is referred to as the \emph{timelock penalty}  The sender further makes use of historical knowledge of channel failures to calculate
\begin{equation}\label{eqn:bias}
    bias_i=\frac{Penalty}{\prod_{j=i}^{n-1}p_j}
\end{equation}
where $Penalty$ is set by default to 100 and $p_j$ is the success probability of the channel $[v_j,v_{j+1}]$. The value of $p_j$ is determined based on the time elapsed since the last failure on channel $[v_j,v_{j+1}]$ according to the sender. More precisely,
\begin{equation}\label{eqn:prob}
    p_j = \begin{cases}
                            P_A\cdot(1-\frac{1}{2^{t/t_{1/2}}})& t  \text{ known}\\
                            P_A                      & t \text{ unknown}\\
                    \end{cases}
\end{equation}
where $P_A$ is a channel-independent apriori probability, $t$ is the time elapsed since the last failure in the channel $(v_j,v_{j+1})$ and $t_{1/2}$, also channel independent, determines the rate with which the probability increases over time. The higher the probability, the more likely it is that the sender adds the channel to the payment path. If the payment fails, the sender is able to track at which channel the failure occurred through error messages. 

\subsection{Collateral Cost}\label{seccost}
If a payment fails at $v_{i+1}$, then the payment amount locked by the intermediary $v_i$ may remain locked for the entire duration of the timelock $T_i$.  The opportunity cost of the locked amount during the duration for which the amount is locked is defined as the \emph{collateral cost}~\cite{miller2017sprites} for the intermediary (in units of \emph{time$\times$ money}). Intuitively, the more coins an intermediary locks, the less balance it has available for other payments, meaning it might have to refuse to lock other payments and lose potential routing fees. Similarly, the longer the funds are locked, the more payments might be missed by the intermediary. Thus, the collateral cost for intermediary $v_i$ is $\mathcal{O}(T_i\cdot \alpha_i)$. 

The routing fee acts as a compensation for the collateral cost. Since the collateral cost $c_i$ is $\mathcal{O}(T_i\cdot \alpha_i)$, we can assume that $c_i$ is directly proportional to $T_i\cdot \alpha_i$. This is similar to the timelock penalty we saw in Section~\ref{seccost}. Thus, we quantitatively define the collateral cost similar to the timelock penalty in Section~\ref{secroute} that represents the cost of locking in \emph{LND's} cost function. The only difference would be that intermediaries would use the value of the total timelock from itself to the receiver instead of the timelock of the channel $(v_i,v_{i+1})$ as used in \emph{LND}. The collateral cost $c_i$ for intermediary $v_i$ is thus $T_i\cdot \alpha_i\cdot r$, where $r$ is the \emph{LND} risk factor.

\subsection{Probing Attacks}\label{secprobing}

Lightning aims to achieve privacy by hiding the balance of a channel from all nodes but the ones adjacent to the channel. In contrast, the capacities of all channels are public information.
Additionally, multi-hop payments in Lightning use onion routing to ensure that any intermediary only knows its predecessor and successor in the path.

\emph{Probing attacks}~\cite{Biryukov2021probing,probing2019herrera,probing2020gijs,probing2022Nisslmueller,probing2020tikhomorov} aim to infer the balances of specific targeted channels.  
If an intermediary does not lock during the first phase of the payment, it is likely that it lacks the necessary balance to complete the payment. By attempting payments with varying payment amounts, the attacker in a probing attack can infer the balance of the channel. Moreover, the attacker ensures that the payment fails, so that it does not have to pay.  Thus, there is no cost associated with the attack but the collateral cost of locking to the intermediaries. 

The balance $B$ can be estimated using a binary search that iteratively finds a smaller interval $[a,b)$ such that the payments succeeds for $a$, i.e., the balance is sufficient, and fails for $b$, i.e., the balance is insufficient. The initial interval is assumed by the attacker to be $[0,C]$, where $C$ is the total capacity of the target channel. The attack only works under the assumption that intermediaries lock whenever they receive a payment they can handle.  An adversary can use the information on channel balances gained through a probing attack to launch Denial-of-Service attacks like \emph{griefing attacks}~\cite{griefing2020zohar,mizrahi2021congestion,perez2020lockdown,rohrer2019discharged,congestion2020Lu}.

The probing attack in~\cite{probing2022Nisslmueller,probing2019herrera} involves the attacker establishing a channel with a target node whose outgoing balances the attacker wants to determine. An attacker sends invalid payments through the target node to the neighbors of the target node to infer the outgoing balances of the target node with its neighbors. 
This attack is improved by utilizing both directions of a channel to probe balances~\cite{probing2020gijs}. Sending  payments over longer paths can be used to probe balances of multiple channels along the path by determining which channel was responsible for the failure~\cite{probing2020tikhomorov}. Finally, recent work in~\cite{Biryukov2021probing} also accounts for multiple channels between two nodes while probing.


\subsection{Detailed Game-Theoretical Background}\label{secgame}


A game here is a situation whose outcome depends on the actions of multiple parties (called players). All players aim to optimize their \emph{utility}, i.e., they assign each outcome a value of how beneficial the outcome is for them and aim to maximize this value. The outcome of the game is then represented as a vector of utilities, one for each player. Note that we use the terms players and parties interchangeably and both mean the same thing. We use Figure~\ref{fig4} to illustrate the concepts introduced in this section. \csg{this seems enough to me}

The outcome of a multi-hop payment depends on the actions of the payer, payee, and intermediaries. Decisions are made sequentially in multi-hop payments, i.e., locking and payment decisions are made one after the other. We can hence use \emph{extensive form games}, which model games as a sequence of decisions, with each sequence of decisions having a particular outcome. Typically, an extensive form game is represented as a \emph{game tree}. The internal nodes of the tree correspond to decisions and are labelled by the name of the player who makes the decision. For instance, in Figure~\ref{fig4}, the root node corresponds to a decision made by $A$ between options $G$ and $H$. $G$ leads to an outcome while $H$ is followed by decisions of other players. 
A link represents an option for a decision leading to a new decision or a final outcome. The link is labelled by the option. Final outcomes, resulting from a sequence of decisions, are represented by leaf nodes. The utilities for every player corresponding to an outcome are represented as a tuple at the corresponding leaf node.

\begin{figure}
	\centering
	\begin{istgame}[font = \tiny,scale = 0.4] 
		\xtdistance{15mm}{30mm} 
		\istroot(0){A}
		\istb{G}[al]{(0,2)}
		\istb{H}[ar]
		\endist
		\istroot(1)(0-2)<above right>{N}
		\istb{S1}[al]
		\istb{S2}[ar]
		\endist
		\xtdistance{15mm}{30mm}
		\istroot(2)(1-1)
		\istb{I}[al]{(2,2)}
		\istb{J}[ar]{(-2,-2)}
		\endist
		\istroot(3)(1-2)
		\istb{I}[al]{(-2,-2)}
		\istb{J}[ar]{(2,2)}
		\endist
		\xtInfoset(2)(3){B}
	\end{istgame}
	\vspace{-0.5em}
	\caption{Illustration of basic game-theoretical concepts }
	
	\label{fig4}
	\vspace{-2.5em}
\end{figure}


Extensive form games differ based on the amount of information the players have. 
The first aspect to consider is whether a player making a decision is aware of all previous decisions. If that is the case for all players, the game has \emph{perfect information}, otherwise it has \emph{imperfect information}. In the game tree, we connect nodes of the same player with a dotted line if the player cannot tell at which node of the game tree the player is making the decision. 

A game has \emph{complete information} if all players know the exact utilities of each outcome and the \emph{type} (e.g., rational, byzantine) of the other players.
Otherwise, the game has \emph{incomplete information}. In a game with incomplete information, there are essentially multiple game trees, one for each possible set of utilities and types. The game can be turned from a game with incomplete information into a game with imperfect information as illustrated in Figure~\ref{fig4}. 
When a decision made by player $P$ has multiple possible outcomes, we add a node with the player called \emph{Nature} to the game tree. Nature represents an external entity that makes random decisions between different scenarios. For each possible outcome, we then add a child labelled $P$ to the Nature's decision.
While the children have the same options available, the options lead to different outcomes. $P$ is unaware which decision \emph{Nature} made and hence does not know in which scenario it is, as in a game with imperfect information.
Figure~\ref{fig4} provides an example. Here, the utilities for $B$ (and $A$) are not fully known if $A$ chooses $H$, they can be either $(2,2)$ for $I$ and $(-2,-2)$ for $J$ if $N$ chooses $S1$ or vice versa if $N$ chooses $S2$ . An external entity \emph{Nature} ($N$) decides whether we are in $S1$ or $S2$.

We now turn to how players make decisions in a game. 
The decisions made by a player throughout the game are the player's \emph{strategy}. One decision is a \emph{move}. A \emph{dominant} strategy is one that achieves a better utility than other alternative strategies, regardless of the actions of the other players.  
A \emph{strategy profile} is the set of strategies used by all the players in the game. 
A \emph{Nash equilibrium} is a strategy profile such that no single player can improve their utility by deviating from their strategy. 

For an extensive form game with perfect and complete information, we can find a Nash equilibrium through \emph{backward induction}. Backward induction starts from the player who represents the leaf node or the last player who chooses the strategy that is most beneficial to itself. The second-to-last player can then predict the last player's strategy (since the game is perfect and has complete information) and can choose its best strategy. In this manner, each player predicts the best strategies of the subsequent players and chooses its best strategy.
The strategy profile corresponding to the one given by decisions chosen during backward induction forms a Nash equilibrium. 

For a game with imperfect information (and hence also for a game with incomplete information), we can use a variant of backward induction that considers expected utilities instead of exact utilities. In order to be able to compute expected utilities, nodes need to have a \emph{belief} about the decisions previously made. Concretely, when uncertain about which node they are at, a player $P$ assigns each potential node a probability, computes the utilities for all of them, and then computes the expected value of these utilities. $P$ chooses the decision with the highest expected utility. 
We can then say that $P$ is acting in a \emph{sequentially rational} manner, \emph{consistent} with their beliefs. The result of all players applying this method is a \emph{Bayesian Nash equilibrium}. 

For instance, in Figure~\ref{fig4}, assume node $B$ assigns probability $q$ to $S1$ and $1-q$ to $S2$ as its belief of $N$'s choice. Then the expected utility for $B$ on choosing $I$ is $2\cdot q -2\cdot (1-q) = 4\cdot q-2$ and similarly the expected utility on choosing $J$ is $2-4\cdot q$. So $B$ chooses $I$ if $q>0.5$ and $J$ if $q<0.5$. 

$A$ in turn has its own belief about what $N$ will choose and also about what $B$ believes (about $q$). Let $q_1$ and $1-q_1$ be the probabilities that $A$ assigns to $S1$ and $S_2$, respectively. Also, let $q_2$ and $1-q_2$ be the probabilities that $A$ assigns to whether $B$ believes $q>0.5$ or $q<0.5$ respectively. After $N$ chooses $S1$, $A$'s expected utility is calculated to be $4\cdot q_2 - 2$.  Similarly, after $N$ chooses $S2$,  $A$'s expected utility is $2-4\cdot q_2$. Thus, $A$'s overall expected utility $u$ on choosing $H$, based on its beliefs, is $q_1\cdot (4\cdot q_2 - 2) + (1-q_1)\cdot(2-4\cdot q_2)$. Since, $A$'s utility on choosing $G$ is $0$, it chooses $H$ only if $u>0$. \csg{I can't find what $u$ means}

Thus far, we have only dealt with \emph{pure} strategies, i.e., each player chooses an action for sure. A player may employ a \emph{mixed} strategy, where it assigns a probability distribution over its set of pure strategies and selects its strategy based on the distribution. For example, in Figure~\ref{fig4}, this means that $A$ chooses $G$ and $H$ with probabilities $p_G$ and $p_H$, respectively ($p_G + p_H=1$).  $A$ assigns $p_G$ (and $p_H$) such that its expected utility $p_G\cdot U_G +p_H\cdot U_H$ is maximized given the values of $q_1$ and $q_2$. Here, $U_G$ and $U_H$ are utilities $A$ obtains by choosing $G$ and $H$, respectively. Similarly, $B$ also assigns a probability distribution over $I$ and $J$ depending on $q$. When each player maximizes their respective utilities, a Bayesian Nash equilibrium is reached.

\section{Analysis of Lightning's Routing}
\label{rout}

Let us now evaluate Lightning's routing algorithm from a game-theoretical perspective. We derive the conditions that incentivize rational intermediaries to lock payments. 

\subsection{Assumptions}
We assume that all parties act rationally and try to maximize their gains from every payment they participate in.  So, we exclude parties that become unresponsive due to being offline. In addition, we consider malicious behavior when we analyze the probing attack by a malicious sender. 
All payments are assumed to be multi-hop and source routed using \emph{LND} protocol. The utilities for intermediaries and senders depend on the routing fee and the collateral cost incurred by locking. 
Since a party sets its base fee and fee rate, it can select their values to be high enough so that fees are always greater than the collateral cost.

Let $(v_0,v_1,\ldots,v_n)$ be a payment in which $v_0$ is sending an amount $\alpha$ to $v_n$, $c_i$ denote $v_i$'s worst-case collateral cost and $T_i=\sum_{j=i}^{n-1}t_i$ denote the cumulative timelock from $v_i$ to $v_n$. In the rest of the paper, we use the terms collateral cost and worst-case collateral cost interchangeably.
The sender initially sends payment value $\alpha_0 = \alpha + \sum_{j=1}^{n-1} f_j$ and each step reduces the payment value by $f_i$, the fees of intermediary $i$. Thus, the amount that $v_{i}$ pays to $v_{i+1}$ in a successful payment is $\alpha_i = \alpha + \sum_{j=i+1}^{n-1} f_j$. In particular, $\alpha_{n-1}=\alpha$.  
The collateral cost $c_i=T_i\cdot \alpha_i\cdot r$ is obtained as described in Section~\ref{seccost}, where $r$ is the \emph{LND} risk factor. 

Since the sender knows the payment path, it can calculate the value of $c_i$ for all intermediaries. The sender $v_0$ is assumed to benefit with a utility gain $U$ when the payment succeeds that more than compensates its loss of funds (i.e., intermediaries' fees and payment amount) and its collateral cost, which is $\alpha\cdot T_0\cdot r$.
Additionally, when a payment fails, the sender knows the exact failure link. This knowledge is useful for the sender to find paths that have a high probability of succeeding and used by Lightning's LND routing algorithm to reduce failures\footnote{\tiny{https://github.com/lightningnetwork/lnd}}.

\begin{figure*}
\centering

\begin{tikzpicture}[font=\tiny,scale=0.75]
    \tikzstyle{solid node}=[circle,draw,inner sep=1.2,fill=black];
    \tikzstyle{hollow node}=[circle,draw,inner sep=1.2];
    \tikzstyle{level 1}=[level distance=10mm]
    \tikzstyle{level 2}=[level distance=13mm]
    \tikzstyle{level 3}=[level distance=16mm]
    \tikzstyle{level 4}=[level distance=19mm]
    \tikzstyle{level 5}=[level distance=22mm]
    \tikzstyle{level 6}=[level distance=25mm]
  \node(0)[hollow node]{}
    child[grow=down]{node[solid node]{}edge from parent node[left]{$NL$}}
    child[grow=right]{node(1)[solid node]{}
      child[grow=down]{node[solid node]{}edge from parent node[left]{$NL$}}
      child[grow=right]{node(2)[solid node]{}
        child[grow=down]{node[solid node]{}edge from parent node[left]{$NL$}}
        child[grow=right]{node(3)[solid node]{}
          child[grow=down]{node[solid node]{}edge from parent node[left]{$DH$}}
          child[grow=right]{node(4)[solid node]{}
            child[grow=down]{node[solid node]{}edge from parent node[left]{$DH$}}
            child[grow=right]{node(5)[solid node]{}
              child[grow=down]{node[solid node]{}edge from parent node[left]{$DH$}}
              child[grow=right]{node(6)[solid node]{}
              edge from parent node[above]{$H$}
              }
            edge from parent node[above]{$H$}
            }
          edge from parent node[above]{$H$}
          }
        edge from parent node[above]{$L$}
        }
      edge from parent node[above]{$L$}
      }
    edge from parent node[above]{$L$}
};

  \node[above]at(0){$v_0$};
  \node[above]at(1){$v_1$};
  \node[above]at(2){$v_2$};
  \node[above]at(3){$v_3$};
  \node[above]at(4){$v_2$};
  \node[above]at(5){$v_1$};
  \node[below]at(0-1){$(0,0,0,0)$};
  \node[below]at(1-1){$(-c_0,0,0,0)$};
  \node[below]at(2-1){$(-c_0,-c_1,0,0)$};
  \node[below]at(3-1){$(-c_0,-c_1,-c_2,0)$};
  \node[below]at(4-1){\makecell{$(-c_0,-c_1,$\\$-c_2-\alpha_2,U)$}};
  \node[below]at(5-1){\makecell{$(-c_0,-c_1-\alpha_1,$\\$f_2-c_2,U)$}};
  \node[right]at(6){\makecell{$(U-c_0 - f_1-f_2-\alpha,$\\ $f_1-c_1,f_2-c_2,U)$}};
\end{tikzpicture}
\vspace{-0.5em}
\caption{Extensive form game for Lightning's original routing protocol with four nodes. Each edge represents one of four moves: lock $L$, not lock $NL$, provide secret $H$, and withhold secret $DH$. The game begins at $v_0$ and leaf nodes  are tuples of utilities. }
\label{figrout}
\vspace{-2em}
   \end{figure*}
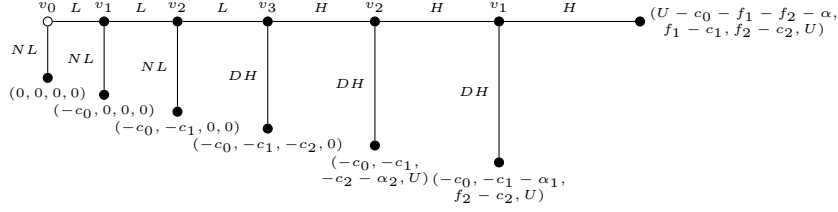

\subsection{Extensive Form Game for Lightning}\label{ext}
The execution of the payment along a path $(v_0,v_1,\ldots,v_n)$ of length $n$ can be considered as a game with $n+1$ players. This execution consists of a locking phase with the $v_i$ and $v_{i+1}$ for $i=0..n-1$ sequentially locking $\alpha_i$ coins in their channel, followed by an unlocking phase in which all locked coins are released starting from the last channel. 
The sender $v_0$ starts the locking phase by making the first move in the game by locking $\alpha_0$ in the channel $(v_0,v_1)$. Each intermediary $v_i$ makes a move after $v_{i-1}$ has locked $\alpha_{i-1}$ in the channel $(v_{i-1},v_i)$. After the last intermediary $v_{n-1}$ locks $\alpha$ in the channel $(v_{n-1},v_n)$, the locking phase of the payment is completed. An intermediary $v_i$ may choose not to lock during this phase either because it has insufficient funds or it is not sufficiently incentivized. In this case, the payment fails with a worst-case utility loss of $c_j$ for each node $v_j$ preceding $v_i$ in the path. The move of locking a payment is denoted by $L$ and the move of not locking is denoted by $NL$.

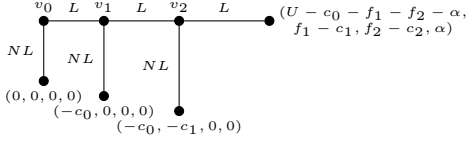
\begin{figure}
 \centering
\begin{tikzpicture}[font=\tiny,scale=1]
    \tikzstyle{solid node}=[circle,draw,inner sep=1.2,fill=black];
    \tikzstyle{hollow node}=[circle,draw,inner sep=1.2];
    \tikzstyle{level 1}=[level distance=8mm]
    \tikzstyle{level 2}=[level distance=10mm]
    \tikzstyle{level 3}=[level distance=12mm]
  \node(0)[solid node]{}
    child[grow=down]{node[solid node]{}edge from parent node[left]{$NL$}}
    child[grow=right]{node(1)[solid node]{}
      child[grow=down]{node[solid node]{}edge from parent node[left]{$NL$}}
      child[grow=right]{node(2)[solid node]{}
        child[grow=down]{node[solid node]{}edge from parent node[left]{$NL$}}
        child[grow=right]{node(3)[solid node]{}
edge from parent node[above]{$L$}
              }
            edge from parent node[above]{$L$}
            }
          edge from parent node[above]{$L$}
          };
       
\node[above]at(0){$v_0$};
  \node[above]at(1){$v_1$};
  \node[above]at(2){$v_2$};
\node[below]at(0-1){$(0,0,0,0)$};
  \node[below]at(1-1){$(-c_0,0,0,0)$};
  \node[below]at(2-1){$(-c_0,-c_1,0,0)$};
   \node[right]at(3){\makecell{$(U-c_0 - f_1-f_2-\alpha,$\\ $f_1-c_1,f_2-c_2,\alpha)$}};
   \end{tikzpicture}
   \vspace{-0.5em}
   \caption{Version of Lightning routing game after concluding that $H$ is always a dominant strategy in comparison to $DH$ and hence players always choose $H$.}
   
   \label{figreduced}
   \vspace{-2em}
   \end{figure}

After all parties have locked, the recipient $v_{n}$ starts the unlocking phase by providing the payment secret to $v_{n-1}$ and hence unlocking the locked amount in the channel $(v_{n-1},v_n)$. Each intermediary $v_i$ again makes a move when it receives the payment secret from $v_{i+1}$. $v_i$ can now share the secret with $v_{i-1}$ to receive the amount locked in the channel $(v_{i-1},v_i)$ or withhold the secret. The move of revealing the secret is denoted by $H$ and the move of withholding the secret is denoted by $DH$. Once $v_i$ receives the locked amount, its net utility in the game is $f_i-c_i$. If $v_i$ withholds the payment secret, it still has to pay $\alpha_i = \alpha + \sum_{j=i+1}^{n-1} f_j$, so its utility is $-c_i - \alpha_i$. The sender does not need to pay in this case as the intermediary does not claim its coins. 
If all intermediaries choose $H$ and $v_0$ receives the secret from $v_1$, $v_0$'s utility is $U-\sum_{i=1}^{n-1}f_i - \alpha -c_0$. An example of the game for $n=3$ is displayed in Figure~\ref{figrout}.   Clearly, after $v_{n-1}$ has locked the payment on its channel $(v_{n-1},v_n)$, all parties are incentivized to choose $H$ over $DH$. As we assume that all intermediaries are rational, they will always choose $H$ and the game tree can be simplified to the one in Figure~\ref{figreduced}.

\begin{figure}
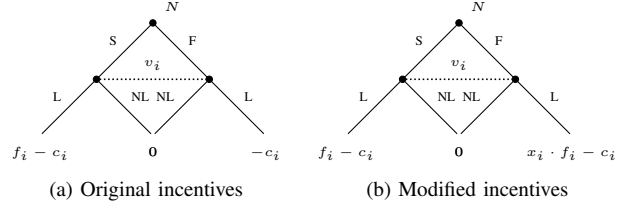

    \centering
\subfloat[Original incentives]{
\begin{istgame}[font=\tiny,scale=0.5] 
\xtdistance{15mm}{30mm} 
\istroot(0)<above right>{$N$}
  \istb{$S$}[al]
  \istb{$F$}[ar]
  \endist
\xtdistance{15mm}{30mm}
\istroot(1)(0-1)
  \istb{$L$}[al]{f_i-c_i}
  \istb{$NL$}[ar]{0}
  \endist
\istroot(2)(0-2)
  \istb{$NL$}[al]{0}
  \istb{$L$}[ar]{-c_i}
  \endist
\xtInfoset(1)(2){$v_i$}
\end{istgame}

\label{figorig}
}
\subfloat[Modified incentives]{
\centering
\begin{istgame}[font=\tiny,scale=0.5] 
\xtdistance{15mm}{30mm} 
\istroot(0)<above right>{$N$}
  \istb{$S$}[al]
  \istb{$F$}[ar]V
  \endist
\xtdistance{15mm}{30mm}
\istroot(1)(0-1)
  \istb{$L$}[al]{f_i-c_i}
  \istb{$NL$}[ar]{0}
  \endist
\istroot(2)(0-2)
  \istb{$NL$}[al]{0}
  \istb{$L$}[ar]{x_i\cdot f_i-c_i}
  \endist
\xtInfoset(1)(2){$v_i$}
\end{istgame}
\label{figmod}
}
\vspace{-0.5em}
\caption{Game tree from the perspective of $v_i$, only showing $v_i$'s utility. $v_i$ cannot be sure whether the payment succeeds ($S$) or fails ($F$) after $v_i$ chooses $L$ ($N$ denotes a random choice here). The differences in the utilities of the original and modified incentives are also illustrated}. 
\vspace{-3em}
\label{figint}

\end{figure}

We now game-theoretically analyze the behavior of payment intermediaries. Both pure and mixed strategies are considered, i.e., intermediaries may always choose the same strategy or choose between multiple strategies. We first assume pure strategies and then generalize. 
Lightning uses onion routing~\cite{goldschlag1999onion}, so an intermediary $v_i$ does not know the complete path of a payment. 
Thus, the Lightning routing game is a game with incomplete information. 
Let $S$ and $F$ be the events indicating that a payment is successful or fails, respectively, after $v_i$ locks the payment. From $v_i$'s perspective, the outcome between $S$ and $F$ is random. However, we assume $v_i$ has a belief on this outcome in the form of probabilities $p^i_S$ and $p^i_F=1-p^i_S$ for success and failure, respectively.
These probabilities are determined by $v_i$ based on the final outcome of past payments it has forwarded to $v_{i+1}$, including the cases where the payment failed due to insufficient balances.
We show the game tree from the perspective of $v_i$ in Figure~\ref{figorig}. 

Let $u_i(L)$ and $u_i(NL)$ denote the utility that $v_i$ expects when choosing $L$ and $NL$ respectively.  
It is clear that $u_i(NL) = 0$. The utility for $v_i$ in choosing $L$ is $f_i-c_i$ if the payment succeeds and $-c_i$ if the payment fails, so $u_i(L) = p^i_S\cdot (f_i-c_i) - p^i_F\cdot c_i$.
Being rational parties, the intermediaries choose $L$ only if the utility for locking is higher than for not locking. Thus, an intermediary $v_i$ forwards the payment only if $u_i(L)>0$. For a successful payment i.e., for locking to be the strategy chosen by all players, we require $u_i(L)>0$ for $i=0,1,\ldots,n$. When each player chooses $L$ only if its beliefs suggest that its utility on choosing $L$ is positive, we have a Bayesian Nash equilibrium. 

Let us now assume that $v_i$ adopts a mixed strategy and assigns probabilities $p_L$ and $p_{NL}$ to the strategies $L$ and $NL$, respectively, with $p_L+p_{NL}=1$. The expected utility of $v_i$ here is $u^m_i=p_L\cdot u_i(L) + p_{NL}\cdot u_i(NL) = p_L\cdot u_i(L)$. We observe that $u^m_i$ is increasing  if $u_i(L)$ is greater than zero and decreasing if it is less than zero. Hence, as long as $u_i(L) \neq 0$, $v_i$ does always adapt a pure strategy, even if mixed strategies are possible. If $u_i(L)=0$, all mixed strategies have the same expected utility, namely 0. 

Since the sender uses the same path-finding protocol as used in \emph{LND}, the exact utility of the sender is not a factor in its decisions. So, we skip the sender's utility computation in this paper.

\section{Routing with Modified Incentives}
\label{sec:modifiedgame}
The current incentive model does not compensate intermediaries for their participation if the payment fails.  If the collateral cost is too high, a rational intermediary may reason to ignore the payment if it believes that the payment is likely to fail.  We propose an alternative model where an intermediary is paid a portion of the fee if it locks the payment amount on its outgoing channel regardless of whether the payment is successful, thus incentivizing intermediaries to lock payments. We analyze the new fee model from a game-theoretical perspective. 

\subsection{Modified Incentive Model}
Let us consider a payment with path $(v_0,v_1,\ldots,v_n)$ of length $n$. In the original model, the intermediary $v_i$ is paid a fee of $f_i$ if the payment succeeds.  We suggest a modification in which the intermediary receives a fee of $x_i\cdot f_i$ if it locks coins regardless of whether the payment succeeds, where $0\leq x_i\leq 1$ is called the \emph{non-refundable fraction}. If the payment succeeds, the remainder of the fee is paid. Here the value of $x_i$ is chosen by the sender $v_0$ and is communicated to the intermediaries during the locking phase together with the payment value $\alpha_i$. Note that the value of $x_i$ can be different for different intermediaries as they may have different collateral costs. The changes to the incentive model only affect the behavior of parties during the locking phase as they are still incentivized to receive funds during the unlocking phase. Thus, the game tree looks identical to Figure~\ref{figorig} but with different utilities. 

Concretely, the intermediary $v_i$ is guaranteed a utility of $x_i\cdot f_i -c_i$ if it chooses to lock during the locking phase when the payment fails and $f_i-c_i$ if the payment succeeds. So, if $v_i$ decides not to lock the payment, then an intermediary $v_j$ that precedes $v_i$ in the payment path has a utility of $x_j\cdot f_j -c_j$ and the sender has a utility of $-c_0-\sum_{j=1}^{i-1}x_j\cdot f_j$. 
Figure~\ref{figmod} displays the game from the perspective of the intermediary $v_i$, analogously to Figure~\ref{figorig}. 


Senders have two options for choosing the value $x_i$. First, they can choose it such that the payment is guaranteed to succeed if balances are sufficient.
However, the fees charged in this case may be higher than necessary. 
Second, the sender can choose the non-refundable fraction $x_i$ such that intermediaries should lock according to the sender's beliefs, leading to lower non-refundable fees than the first option. However, if the sender's beliefs are incorrect, that may lead to failed payments due to the intermediary not having a positive utility according to their beliefs. 
We now discuss the utilities for both cases. For this, let $u^{non}_i$ be the utility function of $v_i$. 

\subsubsection{Case 1: Guaranteed to lock}
Here, the strategy is to set $x_i$ such that $x_i\cdot f_i >c_i$ (note that since $c_i<f_i$, we will find a suitable $x_i$ in $[0,1]$ for all $i$. 
If the value of $x_i\cdot f_i$ is greater than $c_i$ --- or in other words, the utility for $v_i$ on locking $u^{non}_i(L) = x_i\cdot f_i  - c_i>0$ --- then $v_i$ is incentivized to lock regardless of whether the payment succeeds or fails after forwarding.  Thus, $v_i$ will choose $L$ if $x_i\cdot f_i>c_i$.  

\subsubsection{Case 2: Expected to lock} 
Here, the strategy for the sender is to set $x_i$ high enough so that the payment is expected to succeed based on the sender's beliefs, given that there are sufficient funds available. If $x_i\cdot f_i\leq c_i$, then $v_i$ has to consider its belief on $p^i_S$ and $p^i_F$ to calculate its expected utilities on payment success and failure, respectively. As before, $u^{non}_i(NL) = 0$. Its expected utility on choosing $L$ is $u^{non}_i(L) = p^i_S\cdot (f_i - c_i) + p^i_F \cdot (x_i\cdot f_i - c_i)$.
The intermediary $v_i$ hence locks the amount if $u^{non}_i(L)>0$. 
           


While the sender can afford to set $x_i$ as 
in Case 1 because of the high success utility $U$, it can also choose a lower value of $x_i$ (lower than $c_i/f_i$) if it is not willing to lose a large part of its fee. As long as $u^{non}_i(L)>0$ for all intermediaries $v_i$ and channel funds are sufficient, the payment is still expected to succeed.  Based on past observations, the sender has a belief $\Tilde{p}^i_S$ i.e., an estimate of $v_i$'s $p^i_S$. Note that the estimate $\Tilde{p}^i_S$ does not change if $v_i$ refuses to lock as it only is concerned with the success probability after $v_i$ chooses to lock. 

By replacing $p^i_S$ by $\Tilde{p}^i_S$ in our computation $u_i^{non}(L)$, we estimate the value $\Tilde{x}_i$ of $x_i$ for which $u^{non}_i(L)$ equals $0$, i.e., if $x_i = \Tilde{x}_i$, $v_0$'s estimate of $v_i$'s expected utility is zero. Choosing $x_i > \Tilde{x}_i$ is hence necessary for $u^{non}_i(L)>0$, at least according to $v_0$'s belief. However, $v_0$ only is aware of the payments of which it has been the sender. If $v_0$'s belief is not in line with the actual success rate observed by $v_i$, $v_i$ might still not lock and the beliefs need to be updated. 
So, $v_0$ sets the value of $x_i = d_i + \Tilde{x}_i$ or $x_i=1$ if $d_i + \Tilde{x}_i>1$, where $d_i$ is a buffer that increases when $v_i$ does not lock and decreases otherwise, thus the term $d_i$ adapts to new information. We explore ways to choose and adapt $d_i$ in Section~\ref{sec:evaluation}.




In this case, if all intermediaries choose $L$ or $NL$ when their beliefs indicate that $u^{non}_i(L)$ is greater than or less than $0$, respectively,  we have a Bayesian Nash Equilibrium. Similar to the original Lightning protocol, the mixed strategy equilibrium of the routing game is the same as the pure strategy equilibrium.

\subsection{Disincentivizing Probing Attacks}
 In a probing attack (detailed description in Section~\ref{secprobing}), a source uses multiple fake payments to learn the available balance in a channel. 
In the original protocol, the sender does not incur any monetary losses while attempting fake payments. However, with the modified incentives the sender has to pay partial fees whenever the intermediary chooses $L$. Given that a probing attack typically consists of many payments, the intermediary is bound to have a low value for its estimate $p^i_S$ of the probability of a successful payment after locking. Most likely, excluding the first few payments, the source has to choose $x_i\cdot f_i>c_i$ for a rational intermediary to lock. The modified incentives also disincentivize griefing attacks with malicious intermediaries refusing to lock payments as now the non-refundable fees are an opportunity cost for these attackers.

\section{Evaluation}
\label{sec:evaluation}

In this section, we evaluate Lightning's and the modified incentive model in terms of the success ratio of multi-hop payments, the additional income to the intermediaries as well as the increase in cost for the senders. We also experimentally determine the amount of fees that a malicious sender has to pay when conducting a probing attack in our new incentive model. 
\subsection{Simulation setup}\label{simmodel}
Note that we did not specify how to exactly calculate the perceived probabilities of payment success ($p^i_S$,$p'^i_S$ and $P^i_S$ from Sections~\ref{rout},~\ref{sec:modifiedgame}). They are based on past observation but there are multiple ways to weight observations, e.g., based on how in the past they occurred. 
We choose a simple solution for our experiments that is in line with Lightning's general treatment of failures.  
Afterwards, we provide details about how transactions are executed in our setup.

\textbf{Calculation of success probabilities:}
The probabilities $p^i_S$, $P^i_S$ and $\Tilde{p}^i_S$ are calculated in a similar manner as in Eq.~\ref{eqn:prob}. We only consider the last failure instead of, e.g., the frequency of failures, because only the last failure is indicative of the current available funds.
These three probabilities use different values of $t$ and $t_{1/2}$ in Eq.~\ref{eqn:prob}:
an intermediary $v_i$ calculates $p^i_S$ with the value of $t$ being the time since the last payment failed after $v_i$ forwarded a payment to $v_{i+1}$, 
the sender $v_0$ calculates $\Tilde{p}^i_S$ with the value of $t$ being the time 
since the last failure of a payment sent by $v_0$ after $v_i$ locked with $v_{i+1}$, and finally, the sender $v_0$ calculates $P_S^i$ with $t$ being the time since the last failure of a payment sent by $v_0$ caused by $v_i$ refusing to lock with $v_{i+1}$. The sender uses $P_S^i$ to calculate the value of the buffer $d_i$ for $v_i$ with $d_i=0.1\cdot (1-P_S^i)$ so that $d_i$ decrease after $v_i$ locks and increases when $v_i$ does not lock.

We use two different constant values of $t_{1/2}$, one for the intermediaries denoted by $t_{1/2}^I$, and one for senders denoted by $t_{1/2}^S$.  
The value of $t_{1/2}^S$ is always chosen to be higher than $t_{1/2}^I$ to make up for the fact that the sender is unaware of some failures. Hence, it needs to consider the failures it is aware of 
with a higher weight. 
The value of $P_A$ is identical for senders and intermediaries.

\textbf{Payment execution:}
We build on the Python package \emph{Networkx} to simulate a payment channel network and multi-hop payments in the network. 
We did not simulate the cryptographic protocols in order to increase scalability. 
Payments are executed sequentially and assumed to complete --- successfully or not --- immediately.
Having concurrency merely might reduce the success ratio as payments may block each other, but the effect has been shown to be small in comparison to changing the average payment value~\cite{eckey2020splitting}. 
In order to still have a notion of time required for computing the success probabilities, we assume there is a uniformly random delay between two consecutive payments.


Once the sender finds a path based on LND's path finding algorithm, all players make decisions based on their respective expected utilities in the routing games discussed in Section~\ref{rout} and~\ref{sec:modifiedgame}. When a payment fails, the intermediaries who chose $L$ and the sender record this failure and use this knowledge to compute the success probabilities for future payments. If modified incentives are used, then the sender pays partial fees to the intermediaries who have chosen $L$.  

To capture the event where the sender of a payment uses knowledge of past channel failures to adjust the values of $x_i$ for every intermediary, we need multiple payment attempts from the same sender over the same intermediary at different points of time. In our simulations, we obtain such repeated scenarios at a faster rate when we impose a restriction on the set of possible senders and recipients. Hence, we sample a small subset $\omega$ of all the nodes in the network and the sender-recipient pair for every payment are sampled from $\omega$ to save simulation time and yet capture the essential working of the routing game. \csp{New} Meaningful results can be obtained for bigger subsets by simulating a lot more payments. We expect the results to be similar as parties update their strategies only based on the payments they observe.

\subsection{Metrics}\label{metrics}
We use the following metrics to evaluate both the original and modified incentives: 1) The ratio of the number of payments that succeed to the number of payments attempted is the \emph{success ratio} $SR$, 2) The ratio of the number of moves by intermediaries that are $L$ to the number of all the moves made by intermediaries is the \emph{lock ratio} $LR$, 3) The average fee collected by the intermediaries, over all payments that they participate in whether by locking or not locking is called $F_I$ (in satoshis), 4) The average fee paid by the senders,  over all payments they started, is called $F_S$ (in satoshis), 5) The average non-refundable fee gained per intermediary over all failed payments is called $F'_I$(in satoshis), 6) The average non-refundable fee gained per intermediary over all failed payments is called $F'_S$(in satoshis).

\subsection{Parameters}\label{secparameters}

\begin{table*}
	\begin {center}
	\begin{tabular}{c r@{$\pm$}l r@{$\pm$}l r@{$\pm$}l r@{$\pm$}l r@{$\pm$}l r@{$\pm$}l}
		\toprule
		\textit{Cases} 
		& \multicolumn{2}{c}{$SR$} 
		& \multicolumn{2}{c}{$LR$}
		& \multicolumn{2}{c}{$F_I$}
		& \multicolumn{2}{c}{$F_S$}
		& \multicolumn{2}{c}{$F'_I$}
		& \multicolumn{2}{c}{$F'_S$}\\
		\midrule
		\textbf{\textit{Original}} & 0.65&0.11 & 0.70&0.11 & 2.92&0.22&11.26&1.86&0&0&0&0\\
		\textbf{\textit{ModGuaranteed}} & 0.73&0.08 & 0.81&0.07 & 2.67&0.22&10.52&1.85&1.37&0.3&2.63&0.52\\
		\textbf{\textit{ModIncentivized}} & 0.71&0.09 & 0.78&0.08 &2.65&0.27 &10.64&1.94&0.37&0.12&0.73&0.27\\
		\bottomrule
	\end{tabular}
 \vspace{-0.5em}
	\caption{Comparison of metrics in the three cases Original, ModGuaranteed, and ModIncentivized. Each entry comprises of the mean and the standard deviation of the metric averaged across the 10 sets of payments for the corresponding case.}
    \label{table}
	\end {center}
 \vspace{-3em}
\end{table*}

To simulate the payment channel network, we first obtained a snapshot of Lightning from \emph{\url{https://ln.fiatjaf.com}} on February 25th, 2023.  The original snapshot had around 30,000 nodes and 80,000 channels. We 
removed channels that no longer exist and nodes without channels, 
leaving us with around 15,000 nodes and around 60,000 channels.
We also observed that the snapshot included fee (base fee and fee rate) and timelock parameters only for around 8,500 channels.
We derived the distributions of the provided fee and timelock values and then sampled values for the remaining channels independently from the three distributions. 

 The sender and recipient of every payment are sampled randomly from a subset of the network $\omega$ of size 10. 
 Once the sender/recipient pair of a payment has been selected, the payment amount is selected. The main use case for payment channel networks is for payments of smaller values. Small payments are most suitable because blockchain fees for conducting payments on-chain do not depend on the payment value and thus executing smaller payments on-chain are more expensive as compared to larger payments~\footnote{\tiny{https://www.coindesk.com/layer2/paymentsweek/2022/04/26/the-lightning-network-is-bringing-payments-back-to-bitcoin/}}. Keeping this in mind, the payment amount for each payment is sampled uniformly between 5\$  (approximately 16,000 satoshis) and 15\$(approximately 48,000 satoshis). We assume a random delay (chosen uniformly) between 0.1 and 1 minute between consecutive payments.
 
Let us now discuss the selection of the routing game parameters. All parameters discussed here are publicly known to all parties and do not change during the course of a simulation run. Firstly, the value of the \emph{LND} risk factor $r$ is $1.5\cdot 10^{-9}$ by default. The value of $r$ determines how the fee compares with the collateral cost in general and consequently the expected utilitites of senders and intermediaries in payments. We thus determine how the models behave for different values of $r$. The default value used for the probability $P_A$ in LND is $0.6$. 
 The value of the time parameter $t^I_{1/2}$ in the computation of the success probablities is taken to be $30$ minutes, and we set the ratio $\tau = (t_{1/2}^S)/(t^I_{1/2})$ to be $2$.

The following three cases are considered in our evaluations: 1) \textit{Original}: Current Lightning, 2) \textit{ModGuranteed}: The sender incentivizes the intermediary $v_i$ to always lock if they have sufficient funds by setting $x_i=c_i/f_i$ (Case 1 in Section~\ref{sec:modifiedgame}), and 3) \textit{ModIncentivized}: The sender uses beliefs on past failures to set $x_i$ such that the intermediaries $v_i$ are likely to lock (Case 2 in Section~\ref{sec:modifiedgame}).
For each case and parameter setting, we simulate 10 runs of 1000 payments each.

\subsection{Results}\label{results}



We now describe the impact of the various incentives on the success ratio, the lock ratio, as well as the average fees. When the risk factor $r$ was set to $1.5\cdot 10^{-9}$, we observed that the collateral cost was very low in comparison to the fee in nearly 99\% of payments and hence the non-refundable fee was very low. Thus, the case is nearly identical to the original Lightning routing. On the other hand, nearly all payments failed when the risk factor was set to $1.5\cdot 10^{-5}$ as the collateral cost was always higher than the fee. Thus, we only varied the value of $r$ between $1.5\cdot 10^{-6}$, $1.5\cdot 10^{-7}$, and $r=1.5\cdot 10^{-8}$. Here, we focus on the case  $r=1.5\cdot 10^{-7}$.

Table~\ref{table} shows a comparison of the three cases with $P_A = 0.6$ and $\tau=2$. We see that the success ratio of payments increased by around 12.2\% when we used ModGuaranteed instead of Original. The lock ratio also increased similar to the success ratio, namely 15.7\%. We see that the average fee paid by a sender per payment decreased by 6\% and the average fee gained by intermediaries per payment also decreased by 8\%. This decrease was due to the higher number of failed payments for Original, which led to senders choosing paths with higher fees as the bias against paths with recent failures in LND prevents senders from using cheap paths. Even though senders paid partial fees of around $F'_S=2.63$ satoshis  per failed payment when using ModGuaranteed, the difference in the path selection makes up for the extra fees. Thus, at least in the scenarios considered here, having non-refundable fees actually reduced the monetary cost for the sender on average. Intermediaries gained $F'_I=1.37$ satoshis per failed payment, so they benefited from locking as was the goal. The gain was lower than the cost of the sender as the sender typically had to pay multiple intermediaries in a path.


For ModIncentivized, Table~\ref{table} displays that the values of $SR$ and $LR$ increased by 9\% and 11.4\%,  respectively, in comparison to Original.
Although the difference to Original was less pronounced as for ModGuaranteed, ModIncentivized was closer to ModGuaranteed than to Original in terms of locking behavior.  
The results indicate that the sender can choose values of $x_i$ that are lower than $c_i/f_i$ and still expect the payment to succeed. The decrease compared to ModGuaranteed can be attributed to senders not knowing the exact payment history of intermediaries and choosing $x_i$ that were too low to convince the intermediaries to lock. The values of $F_I$ and $F_S$ for ModIncentivized were similar to ModGuaranteed. 
The average income per failed payment for an intermediary was 0.37 satoshis, which is considerably less than for ModGuaranteed, with the sender also paying only about 0.73 satoshis for every failed payment. The difference is to be expected due to the senders choosing lower non-refundable fees.

\begin{figure}
    \centering
    \includegraphics[scale=0.25]{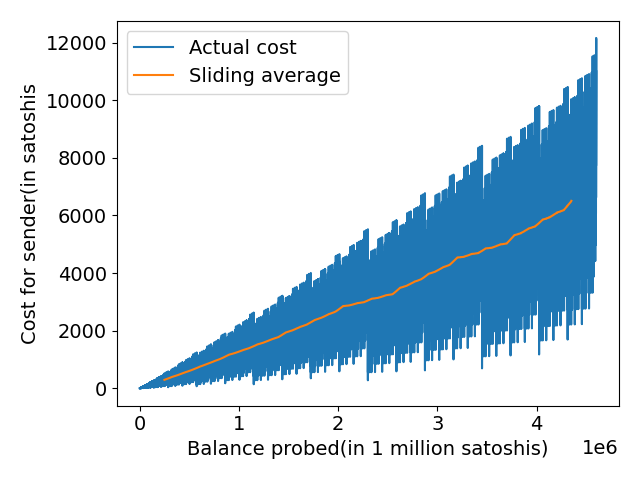}
    \vspace{-1em}
    \caption{The cost for the probing attack for different values of the balance to be inferred in ModGuaranteed. The plot shows both the actual probing cost (in blue) for a particular balance and the sliding average (in orange) for every window of 500,000 satoshis.}
    \label{figprob}
    \vspace{-2em}
\end{figure}

Finally, we evaluated the cost for a malicious sender to conduct a probing attack. 
The attacker followed the ModGuaranteed case for choosing the fee because once the first failure occurs, it had to set the partial fee to $c_i/f_i$ as otherwise, a rational intermediary did not lock. 
Here, we considered a path $(v_0,v_1,v_2)$ where $v_0$ is the malicious sender that tried inferring the forwarding balance that $v_1$ had in its channel with $v_2$. The capacity of the channel $(v_1,v_2)$ was chosen to be 4.6 million satoshis, which is the average channel capacity in Lightning\footnote{\tiny{https://1ml.com/statistics}}. The value of the timelock (or $T_1$) between $v_1$ and $v_2$ was 144 blocks, the default value in Lightning. The balance between $v_1$ and $v_2$ was varied between 0 and the channel capacity and we evaluated the \emph{probing cost} in terms of partial fees needed. The probing attack is essentially a binary search to find the balance, with each search iteration being a fake payment. The probing costs is defined as the total non-refundable fees paid for the fake payments.  

Figure~\ref{figprob} shows the cost for the sender to conduct a probing attack. The total fee paid shows an increasing trend with the balance and the cost is roughly 1500 satoshis for every million satoshis in the balance amount probed. While there is an increasing trend, there is no monotonous increase in cost with the balance. The reason is that one of the factors determining the cost is the number of iterations in which $v_1$ had sufficient balance to lock and was paid a non-refundable fee is not strictly increasing. In addition, the probing cost also depends on the total number of iterations needed for the attack, which is again not strictly increasing.

In summary, we have thus shown that our new incentive model increases the success of payments, as expected. There is no increase in monetary costs for the sender. The new incentive model can also help to deter probing attacks.

\section{Limitations}
In our current simulation setup, we assume certain parameters like the \emph{LND} risk factor $r$ and the apriori success probability $P_A$ are publicly known and identical for all parties. While there are default values in the Lightning implementation, parties may change them to better reflect their individually perceived risk. Parties with a low amount of funds may consider the same collateral cost less acceptable than those that have a lot of liquidity. 

 While we are in line with Lightning's failure treatment in our predictions of success probabilities, their choice is not based on in-depth studies of various methods, which should be done in future work.  Alternatively, other game theoretic methods like \emph{correlated equilibrium}~\footnote{\tiny{https://en.wikipedia.org/wiki/Correlated_equilibrium}} and \emph{auction theory}~\footnote{\tiny{https://en.wikipedia.org/wiki/Auction_theory}} can also be used to analyze Lightning's fee model. While modifying strategies using the success probabilities (as we do) can be considered as a correlated equilibrium, an external party (like a watchtower) can help parties modify their strategies more efficiently to result in a correlated equilibrium at the cost of privacy. Intermediaries can also choose the non-refundable fee in an auction-like manner. However, communication overhead may become an important factor when considering auctions.

Most notably, we did not consider how to realize the non-refundable fees cryptographically, i.e., integrate them into HTLCs. We leave this to future work as it is beyond the scope of our paper.

\section{Related Work}

Several recent works~\cite{avarikioti2019creationgames,Avarikioti_2020ride,rain2021gametheoretic} provide a game-theoretic analysis on different aspects of  payment channel networks. Payment channel networks are treated as creation games in~\cite{avarikioti2019creationgames} and~\cite{Avarikioti_2020ride} where the authors discuss the trade-offs between the creation costs of channels and the fees collected by routing payments. They further analyze various network topologies and determine the fee policies that constitute a Nash equilibrium for each network topology. 
A related work not based on game theory shows how a party can adapt its fee rates to maximize traffic under the assumption that other parties keep their fees as they are~\cite{Ersoy_2020_profit}. However, their results are not applicable when all parties act rational.  

The notion of using extensive form games to analyze payment channel networks to find Bayesian equilibriums was explored in~\cite{rain2021gametheoretic,ZhangGT, mazumdar2022strategic}. The authors in~\cite{rain2021gametheoretic} analyze the security properties of channel closing and multi-hop HTLC payments. They particularly focus on the strategies for both channel parties during the channel closing game and formally characterizing the Wormhole attack~\cite{Malavolta2019AnonymousML} in the routing game. Zhang et.al.~\cite{ZhangGT} analyzes payment channel networks game theoretically in terms of resilience to rational deviations and immunity to Byzantine behaviors. Additionally, the paper also analyzes opening, closing and normal operation of payment channels. However, both works do not consider the negative utilities induced by collateral costs in the routing game. Moreover, they ignore the aspect that intermediaries in payment channel networks do not know the complete payment path and treat the routing game as a game with complete information. \csp{New}The authors in~\cite{mazumdar2022strategic} analyze the intermediaries’ decisions based on the presence of corrupt nodes as a game between two intermediaries. Our model considers all parties involved in the payment including the sender. While we do not model corrupt nodes, we accommodate cases where payments fail due to insufficient balances (which makes identifying corrupt behavior harder).


A parallel work to ours proposes to defend against probing attacks using differential privacy~\cite{van2022hiding}. In contrast to our approach, the defense requires an additional circular noise payment to obfuscate the payment value and hence the inference of the balance. The additional payment requires that more nodes lock collateral, which --- as by our analysis --- some might refuse. 
To date, the work is only the defense against probing attacks.


Like probing attacks, \emph{griefing attacks}~\cite{griefing2020zohar,mizrahi2021congestion,perez2020lockdown,rohrer2019discharged,congestion2020Lu} also exploit the locking mechanism. Instead of privacy, they target the availability of the system by forcing intermediaries to lock funds for as long as possible, preventing the use of these funds in actual payments.
Mazumdar et.al~\cite{counter2020mazumdar,mazumdar2022strategic} propose an alternative HTLC construction that penalizes the node responsible for the griefing attack (intermediary or recipient). Honest victims of the attack are compensated based on its total timelock and the cumulative compensation of all honest victims is paid by the attacker. The alternative HTLC construction involves the creation of two contracts: a penalty contract is set up from the receivers' side whereas the normal payment contract is set up from the sender's side. Thus, it is different to our construction in that our contracts are all set up from the sender side. Furthermore, in our protocol, the sender ---being the attacker in a probing attack --- pays fees for locking whereas in their protocol, the attacker is an intermediary or the receiver and hence the extra fees are not paid by the sender.   

Similar to~\cite{counter2020mazumdar}, the concept of reverse-bond was proposed in~\cite{upfrontreversebond}. The reverse-bond proposal is a mechanism in which a party that accepts a HTLC pays a hold (or rent) fee for the duration that it holds the HTLC. However, there is no proposal on how this can be realized formally and how the hold duration can be tracked accurately. The concept of up-front fees was discussed in~~\cite{nonrefundable,upfront}. However, these proposals were dismissed stating that the solution doesn't avoid the attack and senders might have to pay large upfront fees for small payments.
Several mitigation strategies for griefing like enforcing faster HTLC resolution, loop avoidance, shorter paths and non-refundable fees were suggested in~\cite{mizrahi2021congestion} without any follow-up on how they can be achieved. 


\section{Conclusion}

Lightning's current incentive model falls short of providing incentives to intermediaries when payments fail. We proposed a modified incentives model in which intermediaries are guaranteed partial fees (payed by the sender) for locking payments and analyze both models formally using extensive form games. In the future, we will consider incentives for alternative models of multi-hop payments in payment networks including constant and linear collateral payments~\cite{miller2017sprites,Egger2019amcu,jourenko2021trees}, single-phase payments~\cite{blitz}, and local routing~\cite{speedymurmurs,silentwhispers,robustpay} as well as the design of cryptographic algorithms enforcing advance fees.

\bibliographystyle{plain}
\bibliography{ref}

\begin{thebibliography}{10}

\bibitem{nonrefundable}
Emelyanenkok. 2017. lightning-rfc issue 182: Payment channel congestion via
  spam-attack.
\newblock https://github.com/lightning/bolts/issues/182.

\bibitem{upfront}
A proposal for up-front payments, 2019.
\newblock
  https://lists.linuxfoundation.org/pipermail/lightning-dev/2019-November/002282.html.

\bibitem{upfrontreversebond}
A proposal for up-front payments: Reverse bond payment, 2020.
\newblock
  https://lists.linuxfoundation.org/pipermail/lightning-dev/2020-February/002547.html.

\bibitem{blitz}
Lukas Aumayr, Pedro Moreno-Sanchez, Aniket Kate, and Matteo Maffei.
\newblock Blitz: Secure {Multi-Hop} payments without {Two-Phase} commits.
\newblock In {\em USENIX Security}, 2021.

\bibitem{avarikioti2019creationgames}
Georgia Avarikioti, Rolf Scheuner, and Roger Wattenhofer.
\newblock Payment networks as creation games.
\newblock In {\em Data Privacy Management, Cryptocurrencies and Blockchain
  Technology}, 2019.

\bibitem{Avarikioti_2020ride}
Zeta Avarikioti, Lioba Heimbach, Yuyi Wang, and Roger Wattenhofer.
\newblock Ride the lightning: The game theory of~payment channels.
\newblock In {\em Financial Cryptography and Data Security}. 2020.

\bibitem{Biryukov2021probing}
Alex Biryukov, Gleb Naumenko, and Sergei Tikhomirov.
\newblock Analysis and probing of parallel channels in the lightning network.
\newblock {\em IACR Cryptol. ePrint Arch.}, 2021.

\bibitem{Buterin2013}
Vitalik Buterin.
\newblock Ethereum white paper: A next generation smart contract \&
  decentralized application platform.
\newblock 2013.

\bibitem{decker2015fast}
Christian Decker and Roger Wattenhofer.
\newblock A fast and scalable payment network with bitcoin duplex micropayment
  channels.
\newblock In {\em Symposium on Self-Stabilizing Systems}, 2015.

\bibitem{eckey2020splitting}
Lisa Eckey, Sebastian Faust, Kristina Host{\'a}kov{\'a}, and Stefanie Roos.
\newblock Splitting payments locally while routing interdimensionally.
\newblock {\em Cryptology ePrint Archive}, 2020.

\bibitem{Egger2019amcu}
Christoph Egger, Pedro Moreno-Sanchez, and Matteo Maffei.
\newblock Atomic multi-channel updates with constant collateral in
  bitcoin-compatible payment-channel networks.
\newblock In {\em Proceedings of the 2019 ACM SIGSAC Conference on Computer and
  Communications Security}, 2019.

\bibitem{Ersoy_2020_profit}
O{\u{g}}uzhan Ersoy, Stefanie Roos, and Zekeriya Erkin.
\newblock How to profit from payments channels.
\newblock In {\em Financial Cryptography and Data Security}. Springer
  International Publishing, 2020.

\bibitem{gervais2016security}
Arthur Gervais, Ghassan~O Karame, Karl W{\"u}st, Vasileios Glykantzis, Hubert
  Ritzdorf, and Srdjan Capkun.
\newblock On the security and performance of proof of work blockchains.
\newblock In {\em Proceedings of the 2016 ACM SIGSAC conference on computer and
  communications security}, 2016.

\bibitem{goldschlag1999onion}
David Goldschlag, Michael Reed, and Paul Syverson.
\newblock Onion routing.
\newblock {\em Communications of the ACM}, 1999.

\bibitem{griefing2020zohar}
Jona Harris and Aviv Zohar.
\newblock Flood \& loot: A systemic attack on the lightning network.
\newblock In {\em Proceedings of the 2nd ACM Conference on Advances in
  Financial Technologies}, 2020.

\bibitem{probing2019herrera}
Jordi Herrera-Joancomart\'{\i}, Guillermo Navarro-Arribas, Alejandro
  Ranchal-Pedrosa, Cristina P\'{e}rez-Sol\`{a}, and Joaquin Garcia-Alfaro.
\newblock On the difficulty of hiding the balance of lightning network
  channels.
\newblock In {\em Proceedings of the 2019 ACM Asia Conference on Computer and
  Communications Security}, 2019.

\bibitem{jourenko2021trees}
Maxim Jourenko, Mario Larangeira, and Keisuke Tanaka.
\newblock Payment trees: Low collateral payments for payment channel networks.
\newblock In {\em Financial Cryptography and Data Security}, 2021.

\bibitem{congestion2020Lu}
Zhichun Lu, Runchao Han, and Jiangshan Yu.
\newblock General congestion attack on htlc-based payment channel networks.
\newblock Cryptology ePrint Archive, 2020.

\bibitem{silentwhispers}
Giulio Malavolta, Pedro Moreno-Sanchez, Aniket Kate, and Matteo Maffei.
\newblock Silentwhispers: Enforcing security and privacy in decentralized
  credit networks.
\newblock Cryptology ePrint Archive, 2016.

\bibitem{Malavolta2019AnonymousML}
Giulio Malavolta, Pedro~A. Moreno-Sanchez, Clara Schneidewind, Aniket Kate, and
  Matteo Maffei.
\newblock Anonymous multi-hop locks for blockchain scalability and
  interoperability.
\newblock {\em Proceedings 2019 Network and Distributed System Security
  Symposium}, 2019.

\bibitem{counter2020mazumdar}
Subhra Mazumdar, Prabal Banerjee, and Sushmita Ruj.
\newblock Time is money: Countering griefing attack in lightning network.
\newblock In {\em 2020 IEEE 19th International Conference on Trust, Security
  and Privacy in Computing and Communications (TrustCom)}, 2020.

\bibitem{mazumdar2022strategic}
Subhra Mazumdar, Prabal Banerjee, Abhinandan Sinha, Sushmita Ruj, and Bimal
  Roy.
\newblock Strategic analysis to defend against griefing attack in lightning
  network.
\newblock {\em arXiv preprint arXiv:2203.10533}, 2022.

\bibitem{miller2017sprites}
Andrew~K. Miller, Iddo Bentov, Surya Bakshi, Ranjit Kumaresan, and Patrick
  McCorry.
\newblock Sprites and state channels: Payment networks that go faster than
  lightning.
\newblock In {\em Financial Cryptography and Data Security}, 2017.

\bibitem{mizrahi2021congestion}
Ayelet Mizrahi and Aviv Zohar.
\newblock Congestion attacks in payment channel networks.
\newblock In {\em International Conference on Financial Cryptography and Data
  Security}, 2021.

\bibitem{nakamoto2012bitcoin}
Satoshi Nakamoto.
\newblock Bitcoin: A peer-to-peer electronic cash system, 2009.

\bibitem{probing2022Nisslmueller}
Utz Nisslmueller, Klaus-Tycho Foerster, Stefan Schmid, and Christian Decker.
\newblock Inferring sensitive information in cryptocurrency off-chain networks
  using probing and timing attacks.
\newblock In {\em Information Systems Security and Privacy}, 2022.

\bibitem{perez2020lockdown}
Cristina P{\'e}rez-Sola, Alejandro Ranchal-Pedrosa, Jordi
  Herrera-Joancomart{\'\i}, Guillermo Navarro-Arribas, and Joaquin
  Garcia-Alfaro.
\newblock Lockdown: Balance availability attack against lightning network
  channels.
\newblock In {\em International Conference on Financial Cryptography and Data
  Security}, 2020.

\bibitem{lightning}
Joseph Poon and Thaddeus Dryja.
\newblock The bitcoin lightning network: scalable off-chain instant payments,
  2016.

\bibitem{rain2021gametheoretic}
Sophie Rain, Zeta Avarikioti, Laura Kovács, and Matteo Maffei.
\newblock Towards a game-theoretic security analysis of off-chain protocols,
  2021.

\bibitem{rohrer2019discharged}
Elias Rohrer, Julian Malliaris, and Florian Tschorsch.
\newblock Discharged payment channels: Quantifying the lightning network's
  resilience to topology-based attacks.
\newblock In {\em 2019 IEEE European Symposium on Security and Privacy
  Workshops (EuroS\&PW)}, 2019.

\bibitem{speedymurmurs}
Stefanie Roos, Pedro Moreno-Sanchez, Aniket Kate, and Ian Goldberg.
\newblock Settling payments fast and private: Efficient decentralized routing
  for path-based transactions.
\newblock In {\em NDSS}, 2018.

\bibitem{probing2020tikhomorov}
Sergei Tikhomirov, Rene Pickhardt, Alex Biryukov, and Mariusz Nowostawski.
\newblock Probing channel balances in the lightning network.
\newblock 2020.

\bibitem{van2022hiding}
Gijs van Dam and Rabiah~Abdul Kadir.
\newblock Hiding payments in lightning network with approximate differentially
  private payment channels.
\newblock {\em Computers \& Security}, 2022.

\bibitem{probing2020gijs}
Gijs van Dam, Rabiah~Abdul Kadir, Puteri N.~E. Nohuddin, and Halimah~Badioze
  Zaman.
\newblock Improvements of the balance discovery attack on lightning network
  payment channels.
\newblock In {\em ICT Systems Security and Privacy Protection}, 2020.

\bibitem{ZhangGT}
PeiYun Zhang, ChenXi Li, and MengChu Zhou.
\newblock Game-theoretic modeling and stability analysis of blockchain
  channels.
\newblock In {\em 2020 IEEE International Conference on Systems, Man, and
  Cybernetics}, 2020.

\bibitem{robustpay}
Yuhui Zhang and Dejun Yang.
\newblock Robustpay+: Robust payment routing with approximation guarantee in
  blockchain-based payment channel networks.
\newblock {\em IEEE/ACM Transactions on Networking}, 2021.

\end{thebibliography}

\end{document}